# Modeling formation and transport of clusters at high temperature and pressure gradients by implying partial chemical equilibrium

Eugene V. Stepanov* and Alexander F. Gutsol

*RedShift Energy, Inc., Corpus Christi, TX 78414*.

## Abstract

A comprehensive theory of multi-component molecular transport, based on generalized Fick and Maxwell-Stefan equations, is applied to the transport of clusters in a gas. An approach is developed to describe the transport of an entire ensemble of clusters varying in size as a single species with effective transport coefficients. The major assumption of this approach is the existence of local partial chemical equilibrium between the clusters. It is shown that thermal diffusion emerges in the collective description as a significant factor, even if it is negligible when considering the transport of the original molecular species. Analytical expressions for the effective diffusion and thermal diffusion coefficients under temperature, pressure, and chemical composition gradients are derived. The theory is implemented for a technology of $H_2S$ conversion in a centrifugal plasma-chemical reactor and makes it possible to account for the multitude of sulfur clusters in numerical process modeling.

*) Email: eugene.stepanov@rsenrg.com

## Introduction

An interest in studying molecular clusters, in exploring mechanisms of their formation and a diversity of properties is ongoing [1–3]. Historically, the wide attention developed first upon discovery of fullerenes, then nanotubes, and later many other kinds of clusters with different than carbon compositions were recognized in known processes or synthesized intentionally.

Clusters are molecular formations that are intermediate in between isolated molecules and the classical condensed matter. They are composed from repeated atomic or molecular units thus covering a broad range of sizes. As is different from polymers, the units are not arranged fully or partially into linear chains, and clusters are often treated as compact particles [1–3]. The interest in studying clusters as specific objects constantly grows and expands beyond the original area of the nano-materials onto processes where the cluster formation appears to be important.

One of those areas relates to molecular clusters relevant to processes in atmosphere. The development of clouds is facilitated by the presence of aerosol particles that are considered as nanoclusters. Aerosol-cloud interaction represents one of the largest uncertainties in predicting anthropogenic influence on climate [4,5]. A significant source of particles in atmospheric aerosols is soot, a byproduct of combustion and pyrolysis. Particular chemical composition and morphology of soot may vary as dependent upon the source [6], but a general belief is that soot is formed in combustion by clustering polycyclic aromatic hydrocarbons (PAHs) [7–9]. Clustering has a practical importance in the popular process of hydrogen production by pyrolysis of natural gas in thermal plasma [10,11], where deep understanding of the cluster formation and decomposition can lead to high yield of desirable forms of carbon particles. Studies are also conducted on metal [12–14] and silica [15] clusters for catalysis in chemical reactors, which also aim at experimentally exploring the cluster growth.

Computational modeling of chemical processes is a formidable task that have to combine the complexity of the reaction mechanisms with species transport at conditions of mutually interconnected temperature and pressure distributions [16]. Cluster formation occurs in the environment of chemically reacting molecules, and its modeling implies reproducing concurrent reactions between the molecular species. Even for oxygen-free, thermal pyrolysis of hydrocarbons, the number of species is estimated [17] as of the order of $10^5$, and the number of possible reactions is combinatorial. Obviously, researchers try to decrease the number of reagents to what is deemed sufficient for a particular study [18]. Nonetheless, the demand for computational resources remains immense, and any mathematical simplification that is justified fundamentally would be appreciated.

Another dimension of complexity in modeling arises at conditions of both temperature and pressure gradients when a need to account for transport becomes essential. Heavy clusters are more prone to an action of centrifugal forces, and they may form and decompose as they move between zones with either lower or higher temperatures. It may be noticed that the state of art in the cluster studies is mostly focused on molecular mechanisms with abundant bibliography in references [4–15] cited above. Numerical modeling of the cluster transport can be found only in studies of soot formation in diffusional flames, typical examples of which are presented in Refs. [8,19–22]. However, traditional transport equations of the combustion theory in legacy formulations [23–25] have been used as part of standard FEM (finite element methods) codes developed for the flames. Those equations treat diffusion in multi-component systems by blending rules, and an artificial "correction velocity" is applied in order to compensate for numerical discrepancies in mass fluxes that follow from those rules [26,27]. This may be acceptable if species are diluted in a neutral buffer gas (nitrogen of air) but has to be re-addressed in modeling chemical reactors where reagents can be present in arbitrary concentrations. In addition, treating not only every reagent but also each cluster population as a separate species requires enormous computational resources, which have been resolved in the cited studies approximately by size fractionation.

Alternatively, a comprehensive mathematical theory of multi-component transport on the basis of generalized Fick and Maxwell-Stefan equations [28–30] that stems from the fundamental molecular kinetics theory [31] is available and is continuing to develop [32–37]. This theory has never been applied for the cluster transport. Such application is done in the present study. Furthermore, by utilizing this theory, a theoretical approach is proposed that makes it possible to reduce the description of the transport of the entire multitude of clusters to a transfer of a single, unique species with effective diffusion and thermal diffusion coefficients.

An idea of this approach was disclosed in early publications [38–40] as relevant to diffusional separation of fullerenes in solution. It was demonstrated that the unified description is possible if to assume a partial chemical equilibrium between clusters of different sizes. Concurrently, an effect of thermal diffusion that originated from the temperature dependence of the equilibrium constant was derived. A model was presented in a simplified form, appropriately only to transient transport of dilute clusters in liquid solutions.

The current analysis considers a general case of a mixture of gases with high gradients of parameters such as composition, temperature, and pressure. The goal of this study is to provide closed-form expressions for the effective transport coefficients for both concentration and thermal diffusion of clusters in gases and to explicate a path to use them for overall modeling of processes in chemical systems. A rigorous theory is developed for a ternary mixture, including the cluster-forming species, that can be readily extended to a much larger number of components. For a possible reduction of computational expenses even further, an approximation of the method is proposed and tested in an example of the overall modeling.

In Section 1, a simple case of a diluted cluster-forming species is considered to demonstrate the essence of the approach. Further Sections present a more general theory description.

An application of the theory is probed by modeling the formation and transport of sulfur clusters in a centrifugal plasma-chemical reactor for decomposition of $H_2S$. This is a prospective technology for efficient production of hydrogen that has been described elsewhere [41,42]. The reactor conditions do include high temperature and pressure gradients. By showing the feasibility of the present theory for numerical modeling a comparatively simple chemical system, we believe this may motivate its implementation into FEM codes of much broader use as well as help with interpretation of experimental results in this challenging area of research.

## 1. A demonstration of the approach

Terminology in studies of clusters varies, and at first, we specify terms to use in the current analysis. A cluster is a chemical association of particular atomic or molecular units, and let us call such repeating unit a monomer (as it was called in a past study [15]). A choice of the monomer in clusters may be not unique, so we define a monomer as a molecular formation that is also stable to form vapor. The number of monomers in a cluster is called a cluster number, for which we use indices $n$ or $m$. The species that can form clusters are uniformly denoted as $C$. Other species may exist in gas, and indices to enumerate all species, including $C$, are $\alpha$ and $\beta$. The joint set of species and clusters as separate populations is called components and we use common indices $i, j, k$ for them.

In order to demonstrate a way to use partial chemical equilibrium in a study of transport, let us consider a simple mixture where a dilute species $C$ that is capable of forming clusters is mixed with a buffer gas. Let us assume the gas is subject to parameter gradients, namely a concentration gradient $\mathbf{\nabla}x$, a temperature gradient $\mathbf{\nabla}T$, and a pressure gradient $\mathbf{\nabla}p$. For the concentration, it is convenient to operate with the mole fraction $x$, provided the total volume concentration is $N$, so that the volume mole concentration of the dilute species is $xN$.

The species $C$ are distributed over clusters of different cluster numbers $n$ whose particular mole fractions are $x_n$. The monomer is denoted as $C_1$. The major assumption is that clusters are in a chemical equilibrium between themselves through equilibrium with the monomer (vapor), presumably because of fast attachment-detachment of the monomer by reaction:

$$C_n \leftrightarrow C_1 + C_{n-1} \tag{1.1}$$

with the equilibrium constant $K$ (for mole fractions):

$$K = K(p,T) = \frac{K_p(T)}{p}; \quad K_p(T) = p_0 e^{-\frac{\Delta G}{RT}} \tag{1.2}$$

where $\Delta G$ is the Gibbs energy change for reaction (1.1), $R$ is the gas constant, and $p_0$ is the standard atmospheric pressure. In general, $\Delta G$ can be different for clusters of different sizes, but, at first, we neglect this difference, and account for it later in Section 3.

At the equilibrium, the cluster mole fractions obey relationships that may be obtained iteratively:

$$x_n = \frac{x_1 x_{n-1}}{K}; \quad x_n = x_1 \left(\frac{x_1}{K}\right)^{n-1} = x_1 q^{n-1}; \quad q = \frac{x_1}{K} \tag{1.3}$$

The mole fraction $x_i$ and mass fraction $\omega_i$ of components in a mixture of gases are connected by relationship:

$$\mu\omega_i = \mu_i x_i; \quad \mu = \sum\nolimits_i \mu_i x_i \tag{1.4}$$

where $\mu_i$ is molecular weights of a component and $\mu$ is average molecular weight of the mixture. For the dilute species $C$ that forms clusters, the total mole and mass fractions are calculated as sums of geometric progressions with the common ratio $q$:

$$x = \sum\nolimits_n x_n = x_1 \sum\nolimits_n q^{n-1}; \quad \omega = \sum\nolimits_n \omega_n = \sum\nolimits_n \frac{\mu_n x_n}{\mu} = \frac{\mu_1}{\mu} x_1 \sum\nolimits_n n q^{n-1} \tag{1.5}$$

where $\mu_1$ and $\mu_n$ are molecular weights of a monomer and an *n*-size cluster. As the species $C$ is dilute, $\mu$ is approximately a constant. Presumably, series (1.5) converge, and the upper limit in the sums is formally taken as infinity in this analysis.

In a volume with gradients of the total concentration of species $C$ as well as of temperature $T$ and pressure $p$, the fractional concentrations differ in space, which creates a driving force for diffusion of clusters, including monomer $C_1$. Although each population of clusters diffuses with its own diffusion coefficient, and cluster sizes may span over a very broad range, it appears to be possible to describe the transport of the entire species $C$ with single transport coefficients if the local equilibrium (1.1) is sustained.

To explicate such description, let us consider a local diffusional mass flux $\boldsymbol{j}$ and try to express it in terms of the gas parameter gradients at a given point. In the most general form [28,29], a diffusional flux is calculated by using a diffusional driving force $\boldsymbol{d}_i$ for each component $i$ that for ideal gases with no specific volume forces is defined as follows:

$$\boldsymbol{d}_i = \boldsymbol{\nabla} x_i + (x_i - \omega_i)\boldsymbol{\nabla} \ln p; \quad \sum\nolimits_i \boldsymbol{d}_i = 0 \tag{1.6}$$

where $x_i$ and $\omega_i$ are the mole and mass fractions of the component, respectively. In the case of dilute species, the diffusion coefficient $D_n$ for each cluster population is a binary one for the pair of a given cluster and the solvent gas. The total mass flux by diffusion is a sum of fractional fluxes of all cluster populations that can be expressed [29] by using the driving force as follows:

$$\boldsymbol{j} = \sum\nolimits_n \boldsymbol{j}_n = -\sum\nolimits_n \mu_n N D_n \boldsymbol{d}_n = -\mu_1 N \sum\nolimits_n D_n n[\boldsymbol{\nabla} x_n + (x_n - \omega_n)\boldsymbol{\nabla} \ln p] \tag{1.7}$$

According to relationships (1.3), gradients of cluster mole fractions can be obtained in terms of a gradient of the monomer mole fraction $x_1$:

$$\boldsymbol{\nabla} x_n = n q^{n-1} \boldsymbol{\nabla} x_1 - (n-1) x_1 q^{n-1} \boldsymbol{\nabla} \ln K \tag{1.8}$$

The spatial derivatives of the equilibrium constant $K(p, T)$ can be connected with the gas pressure and temperature gradients by equation (1.2) and by also using a known property [43] of Gibbs free energy $G$:

$$H = -T^2 \left( \frac{\partial}{\partial T} \frac{G}{T} \right)_p \tag{1.9}$$

so that

$$\boldsymbol{\nabla} \ln K = \nu \boldsymbol{\nabla} \ln T - \boldsymbol{\nabla} \ln p; \quad \nu = \frac{\Delta H}{RT} \tag{1.10}$$

where $\Delta H$ is the heat of reaction (1.1) that is positive. Apparently, if the clusters are stable formations, $\nu \gg 1$, which makes thermal diffusion particularly important. The diffusion driving force now is expressed as

$$\boldsymbol{d}_n = \frac{\mu}{\mu_1} \omega_n \boldsymbol{\nabla} \ln x_1 - (n-1) x_n (\nu \boldsymbol{\nabla} \ln T - \boldsymbol{\nabla} \ln p) + (x_n - \omega_n) \boldsymbol{\nabla} \ln p \tag{1.11}$$

Let us introduce a diffusion driving force for the entire species $C$ as an ensemble of clusters:

$$\boldsymbol{d} = \sum\nolimits_n \boldsymbol{d}_n = \boldsymbol{\nabla} x + (x - \omega) \boldsymbol{\nabla} \ln p \tag{1.12}$$

where

$$\boldsymbol{\nabla} x = \sum\nolimits_n \boldsymbol{\nabla} x_n = \frac{\mu}{\mu_1} \omega \boldsymbol{\nabla} \ln x_1 - \left( \frac{\mu}{\mu_1} \omega - x \right) (\nu \boldsymbol{\nabla} \ln T - \boldsymbol{\nabla} \ln p) \tag{1.13}$$

and use $\boldsymbol{\nabla} \ln x_1$ to connect $\boldsymbol{d}_n$ and $\boldsymbol{d}$. After some algebra, one can attain:

$$\boldsymbol{d}_n = \frac{\omega_n}{\omega}\ \boldsymbol{d} + x_n\left(1 - \frac{\mu_1}{\mu_C}n\right)\nu\boldsymbol{\nabla}\ln T \tag{1.14}$$

where $\mu_C = \mu\omega/x$ is average molecular weight of species $C$. By substituting expression (1.14) into (1.7), one may derive the diffusional mass flux of the entire cluster population:

$$\boldsymbol{j} = -D^T\boldsymbol{\nabla}\ln T - \mu_1 ND\boldsymbol{d} \tag{1.15}$$

where the effective transport coefficients:

$$D = \frac{1}{\omega}\sum\nolimits_n D_n n\,\omega_n \tag{1.16}$$

$$D^T = \mu_1\nu N\sum\nolimits_n D_n n x_n\left(1 - \frac{\mu_1}{\mu_C}n\right) \tag{1.17}$$

It may be noticed there is no pressure diffusion term additional to that contained in the driving force $\boldsymbol{d}$. We may conclude that the transport of a species that can form clusters is possible to describe as that of a single species with effective transport coefficients for concentration diffusion and thermal diffusion. This description is instrumental when the transport occurs with concurrent chemical reactions that may supply or destroy the species.

## 2. Generalized Fick equations

A theory of multi-component transport is based on generalized Fick equations that are then transformed to Maxwell-Stefan equations [28–30]. Binary diffusion coefficients in the Maxwell-Stefan equations allow for a molecular interpretation and can be calculated by known empirical correlations or direct measurements. A development of a mathematical model for the transport of clusters with unified transport coefficients that can be explicitly calculated as effective for the entire cluster-forming species is the goal of the present derivation.

Let us consider gas with three species, molecular species *A* and *B*, and species *C* that is present in the form of clusters of presumably unlimited sizes. At the start, the transport of these components is described by a set of an infinite number of generalized Fick equations. We use a set proposed in Ref. [28] with a symmetrical matrix of the Fick diffusion coefficients $\widetilde{D}_{ik} = \widetilde{D}_{ki}$ and a positive sign before terms that contain them:

$$\begin{cases} \boldsymbol{j}_A = \rho\omega_A(\widetilde{D}_{AA}\boldsymbol{d}_A + \widetilde{D}_{AB}\boldsymbol{d}_B + \widetilde{D}_{AC1}\boldsymbol{d}_{C1} + \cdots + \widetilde{D}_{ACn}\boldsymbol{d}_{Cn} + \cdots) \\ \boldsymbol{j}_B = \rho\omega_B(\widetilde{D}_{BA}\boldsymbol{d}_A + \widetilde{D}_{BB}\boldsymbol{d}_B + \widetilde{D}_{BC1}\boldsymbol{d}_{C1} + \cdots + \widetilde{D}_{BCn}\boldsymbol{d}_{Cn} + \cdots) \\ \boldsymbol{j}_{C1} = \rho\omega_{C1}(\widetilde{D}_{C1A}\boldsymbol{d}_A + \widetilde{D}_{C1B}\boldsymbol{d}_B + \widetilde{D}_{C1C1}\boldsymbol{d}_{C1} + \cdots + \widetilde{D}_{C1Cn}\boldsymbol{d}_{Cn} + \cdots) \\ \ldots \\ \boldsymbol{j}_{Cm} = \rho\omega_{Cm}(\widetilde{D}_{CmA}\boldsymbol{d}_A + \widetilde{D}_{CmB}\boldsymbol{d}_B + \widetilde{D}_{CmC1}\boldsymbol{d}_{C1} + \cdots + \widetilde{D}_{CmCn}\boldsymbol{d}_{Cn} + \cdots) \\ \ldots \end{cases} \tag{2.1}$$

Indices are marked by components and, for species $C$, are also enumerated by the cluster numbers, using *n* for cluster numbers in the rows and *m* in the columns. The vector mass fluxes of the components are $\boldsymbol{j}_i$. The diffusion driving forces $\boldsymbol{d}_i$ are defined by equation (1.6). The diffusion coefficients for components in equations (2.1) obey the rule:

$$\sum_k \omega_k \widetilde{D}_{ik} = 0 \tag{2.2}$$

that follows from the definition (1.6). As the next step, let us determine the total mass flux $\boldsymbol{j}_C$ of species $C$ by summation of appropriate rows in set (2.1) with application of rule (2.2):

$$\begin{cases} \boldsymbol{j}_A = \rho\omega_A(\widetilde{D}_{AA}\boldsymbol{d}_A + \widetilde{D}_{AB}\boldsymbol{d}_B + \widetilde{D}_{AC1}\boldsymbol{d}_{C1} + \cdots + \widetilde{D}_{ACn}\boldsymbol{d}_{Cn} + \cdots) \\ \boldsymbol{j}_B = \rho\omega_B(\widetilde{D}_{BA}\boldsymbol{d}_A + \widetilde{D}_{BB}\boldsymbol{d}_B + \widetilde{D}_{BC1}\boldsymbol{d}_{C1} + \cdots + \widetilde{D}_{BCn}\boldsymbol{d}_{Cn} + \cdots) \\ \boldsymbol{j}_C = \rho\omega_C\left(\widetilde{D}_{CA}\boldsymbol{d}_A + \widetilde{D}_{CB}\boldsymbol{d}_B - \dfrac{\omega_A\widetilde{D}_{AC1} + \omega_B\widetilde{D}_{BC1}}{\omega_C}\boldsymbol{d}_{C1} \ldots - \dfrac{\omega_A\widetilde{D}_{ACn} + \omega_B\widetilde{D}_{BCn}}{\omega_C}\boldsymbol{d}_{Cn} \ldots\right) \end{cases} \tag{2.3}$$

where the total mass fraction $\omega_C$ of species $C$ is defined similarly to equation (1.5) by the summation of fractional mass fractions $\omega_{Cn}$ of all clusters:

$$\omega_C = \sum_n \omega_{Cn} = \sum_n \frac{\mu_{Cn}x_{Cn}}{\mu} = \frac{\mu_{C1}}{\mu}\sum_n n x_{Cn} \tag{2.4}$$

with $\mu_{C1}$, $\mu_{Cn}$, and $\mu$ as molecular weights of a monomer, an *n*-size cluster, and average for the total mixture, respectively, and $x_{C1}$ and $x_{Cn}$ as mole fractions of monomers and *n*-size clusters. The effective Fick diffusion coefficients in the equation for $\boldsymbol{j}_C$ can be derived as follows:

$$\widetilde{D}_{CA} = \frac{1}{\omega_C}\sum_m \omega_{Cm}\widetilde{D}_{CmA}\,; \quad \widetilde{D}_{CB} = \frac{1}{\omega_C}\sum_m \omega_{Cm}\widetilde{D}_{CmB} \tag{2.5}$$

In order to perform summation in the rows of the set of equations (2.3), let us split the diffusion driving forces for clusters as is defined in equation (1.6):

$$\boldsymbol{d}_{Cn} = \boldsymbol{\nabla} x_{Cn} + (x_{Cn} - \omega_{Cn})\boldsymbol{\nabla}\ln p \tag{2.6}$$

and define a driving force for the entire species $C$ as follows:

$$\boldsymbol{d}_C = \boldsymbol{\nabla} x_C + (x_C - \omega_C)\boldsymbol{\nabla}\ln p \tag{2.7}$$

where mole fractions $x_{Cn}$ and $x_C$ can be expressed through the mole fraction of monomer $C_1$ by relationships similar to equations (1.3):

$$x_{Cn} = \frac{x_{C1}x_{C(n-1)}}{K}; \quad x_{Cn} = x_{C1}\left(\frac{x_{C1}}{K}\right)^{n-1} = x_{C1}q_C^{n-1}; \quad q_C = \frac{x_{C1}}{K} \tag{2.8}$$

$$x_C = \sum_n x_{Cn} = x_{C1}\sum_n q_C^{n-1}; \quad \omega_C = \sum_n \frac{\mu_{Cn}x_{Cn}}{\mu} = \frac{\mu_{C1}}{\mu}x_{C1}\sum_n n q_C^{n-1} \tag{2.9}$$

The gradients of $x_{Cn}$ and $x_C$ can be related to the monomer gradient $\boldsymbol{\nabla} x_{C1}$. By performing the same algebraic transformations as is done in equations (1.8)-(1.13), one may attain at an expression similar to equation (1.14):

$$\boldsymbol{d}_{Cn} = \frac{\omega_{Cn}}{\omega_C}\boldsymbol{d}_C + x_{Cn}\left(1 - \frac{\mu_{C1}}{\mu_C}n\right)\nu\boldsymbol{\nabla}\ln T \tag{2.10}$$

where $\nu$ is defined by equation (1.10) and $\mu_C = \mu\omega_C/x_C$ is average molecular weight of the entire species $C$.

Now we can perform the summation in the rows of the equation set (2.3) that proceeds by the same algorithm in each row. For components $A$ and $B$ (indexed by $\alpha$) we may write:

$$\sum\nolimits_n \widetilde{D}_{\alpha Cn}\boldsymbol{d}_{Cn} = \boldsymbol{d}_C\frac{1}{\omega_C}\sum\nolimits_n \omega_{Cn}\widetilde{D}_{\alpha Cn} + \nu\boldsymbol{\nabla}\ln T\sum\nolimits_n x_{Cn}\widetilde{D}_{\alpha Cn}\left(1 - \frac{\mu_{C1}}{\mu_C}n\right) \tag{2.11}$$

and summation for species $C$ is expressed as a linear combination of these by equation (2.3). The result of this derivation is the reduction of the infinite set of equations (2.1) to a canonical Fick equation set for a ternary system:

$$\begin{cases} \boldsymbol{j}_A = -D_A^T\boldsymbol{\nabla}\ln T + \omega_A\rho\left(\widetilde{D}_{AA}\boldsymbol{d}_A + \widetilde{D}_{AB}\boldsymbol{d}_B + \widetilde{D}_{AC}\boldsymbol{d}_C\right) \\ \boldsymbol{j}_B = -D_B^T\boldsymbol{\nabla}\ln T + \omega_B\rho\left(\widetilde{D}_{BA}\boldsymbol{d}_A + \widetilde{D}_{BB}\boldsymbol{d}_B + \widetilde{D}_{BC}\boldsymbol{d}_C\right) \\ \boldsymbol{j}_C = -D_C^T\boldsymbol{\nabla}\ln T + \omega_C\rho\left(\widetilde{D}_{CA}\boldsymbol{d}_A + \widetilde{D}_{CB}\boldsymbol{d}_B + \widetilde{D}_{CC}\boldsymbol{d}_C\right) \end{cases} \tag{2.12}$$

where the Fick diffusion coefficient matrix is symmetric, $\widetilde{D}_{AC} = \widetilde{D}_{CA}$ and $\widetilde{D}_{BC} = \widetilde{D}_{CB}$. Because of the symmetry, equations (2.5) can be unified as:

$$\widetilde{D}_{\alpha C} = \frac{1}{\omega_C}\sum\nolimits_n \omega_{Cn}\widetilde{D}_{\alpha Cn} \tag{2.13}$$

where index $\alpha$ stays for $A$ or $B$. The coefficients $\widetilde{D}_{AA}$, $\widetilde{D}_{BB}$, and $\widetilde{D}_{AB} = \widetilde{D}_{BA}$ are the same as in the original set (2.1). The last diagonal element in the diffusion matrix is:

$$\widetilde{D}_{CC} = -\frac{\left(\omega_A\widetilde{D}_{AC} + \omega_B\widetilde{D}_{BC}\right)}{\omega_C} \tag{2.14}$$

The thermal diffusion coefficients are described by the following expressions:

$$D_\alpha^T = -\omega_A\nu\rho\sum\nolimits_n x_{Cn}\widetilde{D}_{\alpha Cn}\left(1 - \frac{\mu_{C1}}{\mu_C}n\right) \tag{2.15}$$

$$D_C^T = -\left(D_A^T + D_B^T\right) \tag{2.16}$$

Similar to the result for a simplified case derived in Section 1, the pressure gradient term is contained in the definitions of the diffusion driving force (2.7).

Apparently, there are no restrictions to extend this analysis to a larger number of molecular gases with which the cluster-forming species $C$ are mixed. For such cases, the index $\alpha$ in equations (2.13) and (2.15) spans over all the included gases.

## 3. An account for clusters of "magic" numbers

It is possible that equilibrium in reaction (1.1) is characterized by equilibrium constants that are different for large and small clusters. Indeed, large clusters essentially are tiny pieces of condensed phase of species $C$. They may be referred as heterophase fluctuations that occur in the vicinity of the condensation line [44]. The equilibrium in this case is more like a phase equilibrium with vapor. Small clusters are often formed by specific molecular interactions [1,2]; they are stronger than the condensed phase and may be composed by monomers in particular numbers that are sometimes called "magic". As an example, sulfur in vapor initially forms [45,46] as $S_2$, $S_4$, $S_6$ and $S_8$. For application to such cases, mathematical expressions for the diffusion coefficients should account for the difference in the equilibrium constants.

Let us denote the maximum "magic" number as $L$, so that, for clusters smaller than $L$, the equilibrium constants for the reaction (1.1) are different and referred as $K_{n-1}$. For clusters larger than $L$, the equilibrium can be approximately considered as being a phase equilibrium with a single equilibrium constant $K$. The mole concentrations of clusters now become:

$$\begin{cases} x_{Cn} = x_{C1}^{n}\left(\prod_{i=1}^{n-1} K_i\right)^{-1}; \quad 1 < n \le L \\ x_{Cn} = x_{CL}q^{n-L}; \quad n > L \end{cases} \tag{3.1}$$

where $q = x_{C1}/K$. The mole and weight fractions of entire species $C$ are defined by summation of expressions (3.1) in a way similar to equations (2.9) but split into two sequences relative to the number $L$ (that are primed and double-primed below):

$$\begin{aligned} x_C &= x_C^{'} + x_C^{''}; \quad \omega_C = \omega_C^{'} + \omega_C^{''} \\ x_C^{'} &= \sum_{n=1}^{L} x_{Cn}; \quad x_C^{''} = \sum_{n=L+1}^{\infty} x_{Cn} = x_{CL}\frac{x_{C1}}{K - x_{C1}} \\ \omega_C^{'} &= \frac{\mu_{C1}}{\mu}\sum_{n=1}^{L} n x_{Cn}; \quad \omega_C^{''} = \frac{\mu_{C1}}{\mu}\sum_{n=L+1}^{\infty} n x_{Cn} = \frac{\mu_{C1}}{\mu} x_{CL} x_{C1}\frac{(L+1)K - Lx_{C1}}{(K - x_{C1})^2} \end{aligned} \tag{3.2}$$

Similar to definition (1.10), we express spatial derivatives of equilibrium constants through gradients of gas temperature and pressure:

$$\nabla \ln K_i = \nu_i \nabla \ln T - \nabla \ln p\,; \quad \nu_i = \left(\frac{\partial \ln K_i}{\partial \ln T}\right)_p = \frac{\Delta H_i}{RT} \tag{3.3}$$

where $\Delta H_i$ are the enthalpy of reaction (1.1) for clusters of particular "magic" number. The diffusion driving forces become:

$$\begin{cases} \boldsymbol{d}_{Cn} = \dfrac{\mu}{\mu_{C1}} \omega \nabla \ln x_{C1} - x_{Cn} \sum_{i=1}^{n-1} \nu_i \nabla \ln T + \omega_{Cn} \left(\dfrac{\mu}{\mu_{C1}} - 1\right) \nabla \ln p; \quad 1 \le n \le L \\ \boldsymbol{d}_{Cn} = \dfrac{\mu}{\mu_{C1}} \omega \nabla \ln x_{C1} - x_{Cn} \left[\sum_{i=1}^{L-1} \nu_i + (n-L)\nu\right] \nabla \ln T + \omega_{Cn} \left(\dfrac{\mu}{\mu_{C1}} - 1\right) \nabla \ln p; \quad n > L \end{cases} \tag{3.4}$$

The total driving force $\boldsymbol{d}_C$ for the entire species $C$ can be obtained by summation of $\boldsymbol{d}_{Cn}$. By eliminating $\nabla \ln x_{C1}$, the sought expression is:

$$\begin{cases} \boldsymbol{d}_{Cn} = \dfrac{\omega_{Cn}}{\omega_C} \boldsymbol{d}_C + \left\{\dfrac{\omega_{Cn}}{\omega_C} \Theta_C - x_{Cn} \sum_{i=1}^{n-1} \nu_i{}' + \nu x_{Cn} \left(1 - \dfrac{\mu_{C1}}{\mu_C} n\right)\right\} \nabla \ln T\,; \quad 1 \le n \le L \\ \boldsymbol{d}_{Cn} = \dfrac{\omega_{Cn}}{\omega_C} \boldsymbol{d}_C + \left\{\dfrac{\omega_{Cn}}{\omega_C} \Theta_C - x_{Cn} \sum_{i=1}^{L-1} \nu_i{}' + \nu x_{Cn} \left(1 - \dfrac{\mu_{C1}}{\mu_C} n\right)\right\} \nabla \ln T; \quad n > L \end{cases} \tag{3.5}$$

where

$$\Theta_C = \sum_{n=1}^{L} x_{Cn} \sum_{i=1}^{n-1} \nu_i{}' + x_C'' \sum_{i=1}^{L-1} \nu_i{}'; \quad \nu_i{}' = \nu_i - \nu \tag{3.6}$$

and $\mu_C = \mu\omega_C / x_C$. The summation in equations (2.3) now should be performed by using equations (3.1) in split intervals below and above $L$. Equations for diffusion coefficients $\widetilde{D}_{AC} = \widetilde{D}_{CA}$ and $\widetilde{D}_{BC} = \widetilde{D}_{CB}$ still preserve their forms (2.13) but the sums are to be taken by using equations (3.1). Equations for thermal diffusion coefficients now account for different $\nu_i$:

$$\begin{aligned} D_\alpha^T = -\omega_\alpha \rho \Bigg\{ & \sum_{n=1}^{\infty} \widetilde{D}_{\alpha Cn} \left[\frac{\omega_{Cn}}{\omega_C} \Theta_C + \nu x_{Cn} \left(1 - \frac{\mu_{C1}}{\mu_C} n\right)\right] \\ & - \sum_{n=1}^{L} \widetilde{D}_{\alpha Cn} x_{Cn} \sum_{i=1}^{n-1} \nu_i{}' - \left(\sum_{i=1}^{L-1} \nu_i{}'\right) \sum_{n=L+1}^{\infty} \widetilde{D}_{\alpha Cn} x_{Cn} \Bigg\} \end{aligned} \tag{3.7}$$

$$D_C^T = -(D_A^T + D_B^T) \tag{3.8}$$

The index $\alpha$ here denotes $A$ or $B$. The thermal diffusivity (3.7) can be also connected with the Fick diffusivity $\widetilde{D}_{\alpha C}$ for the entire species $C$ obtained by equation (2.13):

$$\begin{aligned} D_\alpha^T = -\omega_\alpha \rho \Bigg\{ & \widetilde{D}_{\alpha C} (\Theta_C - \nu x_C) + \sum_{n=1}^{L} \widetilde{D}_{\alpha Cn} x_{Cn} \left(\nu - \sum_{i=1}^{n-1} \nu_i{}'\right) \\ & + \left(\nu - \sum_{i=1}^{L-1} \nu_i{}'\right) \sum_{n=L+1}^{\infty} \widetilde{D}_{\alpha Cn} x_{Cn} \Bigg\} \end{aligned} \tag{3.9}$$

One case remains when all clusters belong to the "magic" subset with $n \le L$. In this case, the fractional driving forces are

$$\boldsymbol{d}_{Cn} = \frac{\omega_{Cn}}{\omega_C}\boldsymbol{d}_C + \left(\frac{\omega_{Cn}}{\omega_C}\sum_{n=1}^{L} x_{Cn}\sum_{i=1}^{n-1}\nu_i - x_{Cn}\sum_{i=1}^{n-1}\nu_i\right)\boldsymbol{\nabla}\ln T \tag{3.10}$$

Accordingly, the thermal diffusion coefficients are:

$$D_\alpha^T = -\omega_\alpha\rho\sum_{n=1}^{L}\widetilde{D}_{\alpha Cn}\left[\frac{\omega_{Cn}}{\omega_C}\sum_{n=1}^{L} x_{Cn}\sum_{i=1}^{n-1}\nu_i - x_{Cn}\sum_{i=1}^{n-1}\nu_i\right] \tag{3.11}$$

$$D_C^T = -(D_A^T + D_B^T) \tag{3.12}$$

## 4. Maxwell-Stefan equations

The equation set (2.12) can be solved analytically for diffusion driving forces $\boldsymbol{d}_\alpha$ provided the mass fluxes $\boldsymbol{j}_\alpha$ are known. The resultant set of equations is called Maxwell-Stefan equations [28–30]:

$$\boldsymbol{d}_\alpha = -\sum_{\beta\neq\alpha}\frac{x_\alpha x_\beta}{D_{\alpha\beta}}\left(\frac{\boldsymbol{j}_\alpha}{\rho_\alpha} - \frac{\boldsymbol{j}_\beta}{\rho_\beta}\right) - \sum_{\beta\neq\alpha}\frac{x_\alpha x_\beta}{D_{\alpha\beta}}\left(\frac{D_\alpha^T}{\rho_\alpha} - \frac{D_\beta^T}{\rho_\beta}\right)\boldsymbol{\nabla}\ln T \tag{4.1}$$

where indices $\alpha$ and $\beta$ refer to diffusing species, $A$, $B$, and $C$. The species densities are $\rho_\alpha = \omega_\alpha\rho$ (or $\rho_\beta = \omega_\beta\rho$). The coefficients $D_{\alpha\beta}$ are binary Maxwell-Stefan diffusivities, parameters that are believed to be subject to molecular interpretation and empirical correlation, contrary to the generalized Fick diffusion coefficients in equations (2.12). Another advantage is that diffusional mass fluxes $\boldsymbol{j}_\alpha$ participate in equation (4.1) as differences and can be replaced [28] by total mass fluxes $\boldsymbol{J}_\alpha$, diffusional plus convectional ones, which is important if chemical kinetics is modeled concurrently with transport. However, the partial equilibrium method for transport has been developed in Sections 2 and 3 just for the Fick diffusion coefficients $\widetilde{D}_{\alpha\beta}$, and the method has to include a procedure to connect them with the binary diffusivities $D_{\alpha\beta}$.

Rigorous mathematics of this connection would first require building a matrix of binary diffusivities $D_{ij}$ for all components, including clusters, and knowing the component mole and mass fractions. Then the matrix $D_{ij}$ has to be transformed and appropriately inverted in order to fit the structure of the Fick equations (2.1). As the matrix of Fick diffusivities $\widetilde{D}_{ij}$ is calculated, coefficients $\widetilde{D}_{\alpha Cn}$ become known and equations of Sections 2 or 3 can be applied. The obtained matrix $\widetilde{D}_{\alpha\beta}$ of a reduced rank that contains entire species only has to be converted back to the binary coefficients $D_{\alpha\beta}$ in order to be used in equations (4.1). The technique of the mutual transformation between $D_{ij}$ and $\widetilde{D}_{ij}$ has been developed and is available in literature [29,30,47]. However, it varies in details in order to pursue objectives of particular studies and no one is suited completely for the present analysis. On one hand, the technique should be applicable to the equation set (2.1) with symmetric matrix $\widetilde{D}_{ij}$, and on the other hand, the numerical algorithm that implements it has to be stable in a broad range of cluster concentrations. We briefly revisit this technique to formulate it with a focus on a possibility of very small concentrations of the clusters so that they should appear in no more than the first power at each step of numerical computations.

For the diffusion coefficients, we use reduced quantities:

$$\mathcal{D}_{ij} = N D_{ij}; \quad \widetilde{\mathcal{D}}_{ij} = N \widetilde{D}_{ij} \tag{4.2}$$

where $N$ is the mole concentration of the gas mixture with the average molecular weight $\mu$, so that $\rho = \mu N$. The reduced diffusivities (4.2) depend upon temperature only. Then, similar to a path of derivation that was developed in Ref. [47], we define quantity

$$\Lambda_{ij} = \frac{1}{\mu \mathcal{D}_{ij}} \frac{x_i x_j}{\omega_i \omega_j} = \frac{1}{\mathcal{D}_{ij}} \frac{\mu}{\mu_i \mu_j} \tag{4.3}$$

that does not depend on the component concentrations. Maxwell-Stefan equations for all components that are connected with the original set of Fick equations (2.1) do not have thermal diffusion terms. In the reduced quantities, they are written as:

$$\boldsymbol{d}_i = \sum\nolimits_{j \neq i} \Lambda_{ij} \left( \omega_i \boldsymbol{j}_j - \omega_j \boldsymbol{j}_i \right) \tag{4.4}$$

Matrix $\Lambda_{ij}$ has no diagonal elements defined yet. The path [47] is to define them as

$$\omega_i \Lambda_{ii} = -\sum\nolimits_{j \neq i} \Lambda_{ij} \omega_j \tag{4.5}$$

which transforms equations (4.4) into a form suitable for matrix operations:

$$\boldsymbol{d}_i = \omega_i \sum\nolimits_j \Lambda_{ij} \boldsymbol{j}_j \tag{4.6}$$

Now, if we express the Fick equations (2.1) as

$$\boldsymbol{j}_i = \mu \omega_i \sum\nolimits_k \widetilde{\mathcal{D}}_{ik} \boldsymbol{d}_k \tag{4.7}$$

we attain at the sought connection:

$$\boldsymbol{d}_i = \mu \sum\nolimits_{j,k} \omega_i \Lambda_{ij} \left[ \omega_j \widetilde{\mathcal{D}}_{jk} \right] \boldsymbol{d}_k \tag{4.8}$$

that determines identity:

$$\mu \sum_j \omega_i \Lambda_{ij} \left[ \omega_j \widetilde{\mathcal{D}}_{jk} \right] = \delta_{ik} - \omega_i \tag{4.9}$$

One may notice this equation includes term $\omega_i$ that is constant over subset $k$ in the right side of equation (4.8) and seemingly may be replaced by any other constant because the sum of components of $\boldsymbol{d}_k$ is zero by equation (1.6). It has been argued (Ref. [30], section 4) that identity (4.9) has to hold when multiplied by $\omega_k$ and summed over $k$, which yields just the term $\omega_i$.

Equation (4.9) can be written in a matrix form:

$$\Psi \Omega \widetilde{\mathcal{D}} = Y \tag{4.10}$$

$$\Psi_{ij} = \mu \omega_i \Lambda_{ij}; \quad \Omega_{ij} = \omega_i \delta_{ij}; \quad Y_{ij} = \delta_{ij} - \omega_i \tag{4.11}$$

Because of definition (4.5), the matrix $\Psi$ is singular. The solution is achieved by forming a non-singular matrix $\Psi^0$ where elements of $\Psi$ are subtracted by its diagonal elements in each row:

$$\Psi_{ij}^0 = \Psi_{ij} - \Psi_{ii} = \mu \left( \omega_i \Lambda_{ij} + \sum_{k \neq i} \Lambda_{ik} \omega_k \right) \tag{4.12}$$

The property (2.2) makes sure that equation (4.10) does not change upon replacing $\Psi$ by $\Psi^0$. Finally, the Fick diffusion coefficients can be obtained numerically by inverting $\Psi^0$ and then dividing by elements of the diagonal matrix $\Omega$:

$$\widetilde{\mathcal{D}} = \Omega^{-1} (\Psi^0)^{-1} Y \tag{4.13}$$

A subset of the coefficients $\widetilde{\mathcal{D}}_{ij}$ corresponds to the diffusion of clusters of species $C$ as described in the original set of equations (2.1). They are then used in calculations of the effective coefficients of both concentration diffusion $\widetilde{\mathcal{D}}_{\alpha C}$ and thermal diffusion $D_\alpha^T$ for the entire species that are explicated in Sections 2 and 3. The last step in this derivation is the conversion of the effective transport coefficients back to the binary Maxwell-Stefan diffusivities. For the ternary system of species $A$, $B$, and $C$, equations for this conversion are available in the literature (Ref. [28], Table 24.2.2), which we adapt by utilizing relationships (1.4) and (4.2):

$$\mu_A\mu_B\mathcal{D}_{AB} = \mu^2\frac{\widetilde{\mathcal{D}}_{AB}\widetilde{\mathcal{D}}_{CC} - \widetilde{\mathcal{D}}_{AC}\widetilde{\mathcal{D}}_{BC}}{\widetilde{\mathcal{D}}_{AB} + \widetilde{\mathcal{D}}_{CC} - \widetilde{\mathcal{D}}_{AC} - \widetilde{\mathcal{D}}_{BC}}$$
$$\mu_B\mu_C\mathcal{D}_{BC} = \mu^2\frac{\widetilde{\mathcal{D}}_{BC}\widetilde{\mathcal{D}}_{AA} - \widetilde{\mathcal{D}}_{AB}\widetilde{\mathcal{D}}_{AC}}{\widetilde{\mathcal{D}}_{BC} + \widetilde{\mathcal{D}}_{AA} - \widetilde{\mathcal{D}}_{AB} - \widetilde{\mathcal{D}}_{AC}} \qquad (4.14)$$
$$\mu_A\mu_C\mathcal{D}_{AC} = \mu^2\frac{\widetilde{\mathcal{D}}_{AC}\widetilde{\mathcal{D}}_{BB} - \widetilde{\mathcal{D}}_{BC}\widetilde{\mathcal{D}}_{AB}}{\widetilde{\mathcal{D}}_{AC} + \widetilde{\mathcal{D}}_{BB} - \widetilde{\mathcal{D}}_{BC} - \widetilde{\mathcal{D}}_{AB}}$$

(Entries in table (4.14) mutually correlate by cyclic permutation of indices A-B-C-A… and by also taking into account that matrix $\widetilde{\mathcal{D}}$ is symmetric). Equations (4.1) are slightly modified to utilize quantities (4.14) as follows:

$$\boldsymbol{d}_\alpha = -\mu\sum_{\beta\neq\alpha}\frac{\omega_\beta\boldsymbol{j}_\alpha - \omega_\alpha\boldsymbol{j}_\beta}{\mu_\alpha\mu_\beta\mathcal{D}_{\alpha\beta}} - \mu\sum_{\beta\neq\alpha}\frac{\omega_\beta D_\alpha^T - \omega_\alpha D_\beta^T}{\mu_\alpha\mu_\beta\mathcal{D}_{\alpha\beta}}\boldsymbol{\nabla}\ln T \qquad (4.15)$$

Indices $\alpha$ and $\beta$ here refer to all of the diffusing species $A$, $B$, and $C$.

Obviously, this direct matrix procedure as a whole can be performed only on a limited set of clusters. The size of this set is actually determined by computational resources, particularly taking into account that the matrix inversion (4.13) has to be made in each cell and in each computer iteration. A desirable expansion of this set motivates seeking an approximate procedure that would still provide reasonable results. As such, we propose to use an approximation of a dilute mixture that stems from a fact that the larger a cluster is, the smaller its mole fraction is. In this case, for the purpose of the conversion of fractional diffusivities, we may treat the cluster populations as independent species that propagate in a volume of a buffer gas (that has the average molecular weight $\mu$ of the mixture). For a system of only two components, $S$ as the species and $P$ as the buffer, it is known that (Ref. [28], Table 24.2.1):

$$\widetilde{D}_{SP} = \frac{\omega_S\omega_P}{x_S x_P}D_{SP} \cong \frac{\mu_S}{\mu}D_{SP}; \quad \text{if } x_S \ll 1, x_P \cong 1 \qquad (4.16)$$

that leads to an approximation for the conversion $D_{ij}$ to $\widetilde{D}_{ij}$ and back, $\widetilde{D}_{\alpha\beta}$ to $D_{\alpha\beta}$, as follows:

$$\widetilde{D}_{\alpha Cn} \cong \frac{\mu_{C1}}{\mu}nD_{\alpha Cn}; \quad D_{\alpha C} \cong \frac{\mu}{\mu_C}\widetilde{D}_{\alpha C} \qquad (4.17)$$

where $\mu_{C1}$ is the molecular weight of a monomer and $\mu_C$ is average molecular weight for the entire cluster-forming species $C$, $\mu_C = \mu\omega_C/x_C$.

By comparing equations (2.13), (2.15) with equations (1.16), (1.17) one may see that the substitution (4.17) does correspond to the accurate calculations for a dilute substance. However, we realize the approximation (4.17) is heuristic for a multi-component mixture, and we are to verify that it holds in a practical application by comparing results of approximate and rigorous computations in the next section.

Explicit forms for the transport coefficients to use in equations (4.15) for the approximate solution are presented in Table 1. As an example of cluster populations, relevant to the trial computations in the next section, we consider a set with “magic” numbers 2,3, and 4, and a continuum of cluster sizes above. The transport coefficients are derived from equations (2.13) and (3.11) with rule (4.17) for systems with $n \leq L$ when only clusters of “magic” numbers are present. In consecutive rows of Table 1, $L$ is assumed to increase from 2 to 4. When all sizes of clusters above 4 are taken into account, equation (3.11) is replaced by equation (3.7) in the derivation. The last row in Table 1 lists the expressions for the latter, most general case.

Table 1: Binary transport coefficients in the approximate model with an account for clusters of "magic" numbers up to size $L \leq 4$

| Equilibria | Symbols related | Equations (see Section 3 for definitions) |
|---|---|---|
| $C_2 \leftrightarrow 2C_1$ | $K_1, \nu_1$ | $x_{C2} = x_{C1}\dfrac{x_{C1}}{K_1}; \quad \omega_{C2} = 2\dfrac{\mu_{C1}}{\mu}x_{C2}$<br>$\mathcal{D}_{\alpha C} = \dfrac{\mu_{C1}}{\mu_C}\dfrac{1}{\omega_C}(\omega_{C1}\mathcal{D}_{\alpha C1} + 2\omega_{C2}\mathcal{D}_{\alpha C2})$<br>$z = x_{C2}\nu_1$<br>$D_\alpha^T = -\omega_\alpha(z\mathcal{D}_{\alpha C}\mu_C - 2\mu_{C1}x_{C2}\nu_1\mathcal{D}_{\alpha C2})$ |
| $C_2 \leftrightarrow 2C_1$<br>$C_3 \leftrightarrow C_1 + C_2$ | $K_1, \nu_1$<br>$K_2, \nu_2$ | $x_{C2} = x_{C1}\dfrac{x_{C1}}{K_1}; \quad x_{C3} = x_{C1}\dfrac{x_{C1}^2}{K_1 K_2}; \quad \omega_{Cn} = \dfrac{\mu_{C1}}{\mu}n x_{Cn}$<br>$\mathcal{D}_{\alpha C} = \dfrac{\mu_{C1}}{\mu_C}\dfrac{1}{\omega_C}(\omega_{C1}\mathcal{D}_{\alpha C1} + 2\omega_{C2}\mathcal{D}_{\alpha C2} + 3\omega_{C3}\mathcal{D}_{\alpha C3})$<br>$z = x_{C2}\nu_1 + x_{C3}(\nu_1 + \nu_2)$<br>$D_\alpha^T = -\omega_\alpha\{z\mathcal{D}_{\alpha C}\mu_C - \mu_{C1}[2x_{C2}\nu_1\mathcal{D}_{\alpha C2} + 3x_{C3}(\nu_1 + \nu_2)\mathcal{D}_{\alpha C3}]\}$ |
| $C_2 \leftrightarrow 2C_1$<br>$C_3 \leftrightarrow C_1 + C_2$<br>$C_4 \leftrightarrow C_1 + C_3$ | $K_1, \nu_1$<br>$K_2, \nu_2$<br>$K_3, \nu_3$ | $x_{C2} = x_{C1}\dfrac{x_{C1}}{K_1}; \quad x_{C3} = x_{C1}\dfrac{x_{C1}^2}{K_1 K_2}; \quad x_{C4} = x_{C1}\dfrac{x_{C1}^3}{K_1 K_2 K_3}; \quad \omega_{Cn} = \dfrac{\mu_{C1}}{\mu}n x_{Cn}$<br>$\mathcal{D}_{\alpha C} = \dfrac{\mu_{C1}}{\mu_C}\dfrac{1}{\omega_C}(\omega_{C1}\mathcal{D}_{\alpha C1} + 2\omega_{C2}\mathcal{D}_{\alpha C2} + 3\omega_{C3}\mathcal{D}_{\alpha C3} + 4\omega_{C4}\mathcal{D}_{\alpha C4})$<br>$z = x_{C2}\nu_1 + x_{C3}(\nu_1 + \nu_2) + x_{C4}(\nu_1 + \nu_2 + \nu_3)$<br>$D_\alpha^T = -\omega_\alpha\{z\mathcal{D}_{\alpha C}\mu_C$<br>$\quad - \mu_{C1}[2x_{C2}\nu_1\mathcal{D}_{\alpha C2} + 3x_{C3}(\nu_1 + \nu_2)\mathcal{D}_{\alpha C3} + 4x_{C4}(\nu_1 + \nu_2 + \nu_3)\mathcal{D}_{\alpha C4}]\}$ |

Table 1 (continuing)

| Equilibria | Symbols related | Equations (see Section 3 for definitions) |
|---|---|---|
| $C_2 \leftrightarrow 2C_1$<br>$C_3 \leftrightarrow C_1 + C_2$<br>$C_4 \leftrightarrow C_1 + C_3$<br>$C_5 \leftrightarrow C_1 + C_4$<br>…<br>$C_n \leftrightarrow C_1 + C_{n-1}$<br>… | $K_1, \nu_1, \nu_1^{'} = \nu_1 - \nu$<br>$K_2, \nu_2, \nu_2^{'} = \nu_2 - \nu$<br>$K_3, \nu_3, \nu_3^{'} = \nu_3 - \nu$<br>$K, \nu, \nu_4^{'} = 0$<br>…<br>$K, \nu, \nu_n^{'} = 0$<br>… | $x_{C2} = x_{C1}\frac{x_{C1}}{K_1}; \quad x_{C3} = x_{C1}\frac{x_{C1}^2}{K_1 K_2}; \quad x_{C4} = x_{C1}\frac{x_{C1}^3}{K_1 K_2 K_3}; \quad \omega_{Cn} = \frac{\mu_{C1}}{\mu} n x_{Cn}$<br>$\mathcal{D}_{\alpha C} = \frac{\mu_{C1}}{\mu_C}\frac{1}{\omega_C}\left(\omega_{C1}\mathcal{D}_{\alpha C1} + 2\omega_{C2}\mathcal{D}_{\alpha C2} + 3\omega_{C3}\mathcal{D}_{\alpha C3} + 4\omega_{C4}\mathcal{D}_{\alpha C4} + \sum_{n=5}^{\infty} n\mathcal{D}_{\alpha Cn}\omega_{Cn}\right)$<br>$z = x_{C2}\nu_1^{'} + x_{C3}(\nu_1^{'} + \nu_2^{'}) + (x_{C4} + x_C^{''})(\nu_1^{'} + \nu_2^{'} + \nu_3^{'}) - \nu x_C$<br>$D_\alpha^T = -\omega_\alpha \Big\{ z\mathcal{D}_{\alpha C}\mu_C$<br>$+ \mu_{C1}\Big[\nu x_{C1}\mathcal{D}_{\alpha C1} + 2x_{C2}(\nu - \nu_1^{'})\mathcal{D}_{\alpha C2} + 3x_{C3}(\nu - \nu_1^{'} - \nu_2^{'})\mathcal{D}_{\alpha C3}$<br>$+ (\nu - \nu_1^{'} - \nu_2^{'} - \nu_3^{'})\left(4x_{C4}\mathcal{D}_{\alpha C4} + \sum_{n=5}^{\infty} n\mathcal{D}_{\alpha Cn}x_{Cn}\right)\Big]\Big\}$ |

## 5. Discussion

There are many examples of gas flows with high concentration, temperature, and pressure gradients in chemical engineering. To probe the developed theory, one may choose a model process system where all of them are present and prominent in the same volume, and formation of clusters of at least one of reacting species is proven. In this quality, we consider a promising technology for producing hydrogen by thermal decomposition of $H_2S$ in a centrifugal plasma-chemical reactor [41,42]. An existence of a long line of sulfur clusters in a gas phase has been explored rather well [45,46].

The centrifugal reactor is of "tornado-type" where the gas flow is injected tangentially into a cylindrical vessel with almost sonic velocity and then circulates rotationally spiraling along the axis of the vessel. A concurrent slower radial flow toward the axis of rotation consumes the gas into an axial counterflow for the gas to leave the system through an axial opening [48,49]. A powerful electric discharge is positioned along the axis and causes thermal decomposition. The overall chemical reaction in gas is endothermal and is outlined as follows:

$$H_2S \rightleftarrows H_2 + \frac{1}{2}S_2 \tag{5.1}$$

which indicates that sulfur is originally produced as $S_2$ before conversion into clusters and eventually to condensed state [50]. Furthermore, most stable clusters have the sulfur numbers as multipliers of 2, i.e. $S_4$, $S_6$, and $S_8$, and such trend may be seen also in larger clusters [45,46].

Thus, it is reasonable to assume that the cluster monomer of the present theory is molecule $S_2$ with which the reaction (1.1) should be written. In this case, the number of sulfur atoms in a cluster is twice the number $n$ of monomer units in that cluster (the cluster number).

It is believed that salient features of the gas flow in the centrifugal plasma reactor can be effectively reproduced by a computational 1D model. A schematic and major equations of the model are outlined in Appendix. While the reactor geometry is simplified, the model delivers a detailed description of both kinetic and transport processes and provides concentration profiles for species as well as distributions of the gas parameters in the volume. The present theory for clusters has been fully incorporated into the computation with inclusion of "magic" clusters $S_4$, $S_6$, and $S_8$ for which the profiles are generated separately. The theory is used in two versions as is presented in Section 4, a direct (matrix) version and an approximate one.

While the direct version is mathematically strict, the approximate one provides much larger coverage of cluster sizes, and identifying process conditions under which restrictions on the

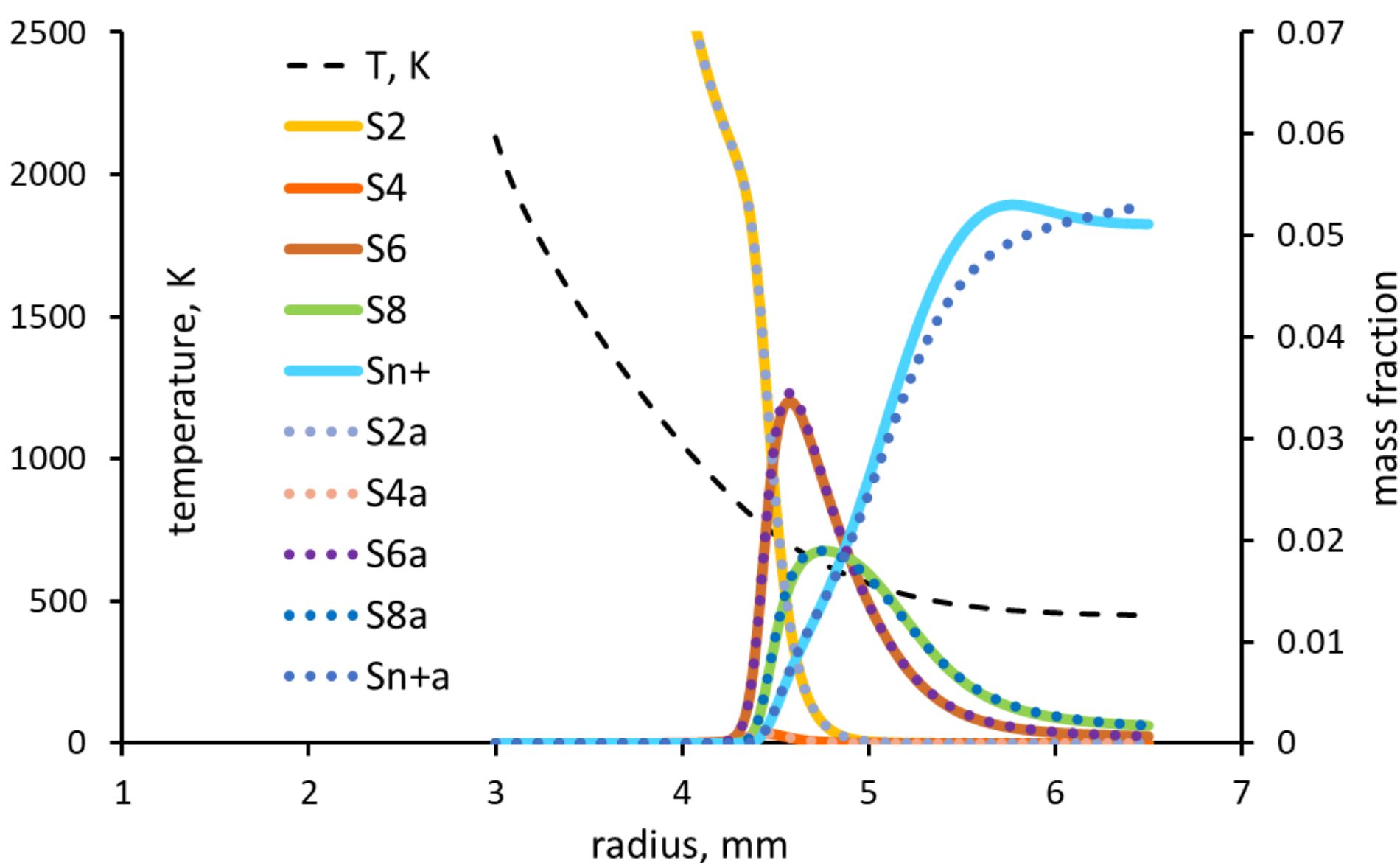

Figure 1. Comparison of spatial distributions of sulfur clusters in a centrifugal plasma-chemical reactor predicted by the direct model and by the approximate one. The direct model operates with 36 cluster numbers (up to $S_{72}$). Curves computed with the approximate model are marked by suffix "a" in the legend. Sn+ stands for cumulative mass fraction of clusters $S_{10}$ and above.

---

cluster sizes included into computations become important has been one of the objectives of the study.

In most simulations performed the direct and approximate versions produced sufficiently close results. Concentration profiles of sulfur clusters at typical conditions in the reactor when the $H_2S$ gas is injected at atmospheric pressure are presented in Figure 1. Computations with the direct model have been done with matrices of 36 clusters, i.e. up to $S_{72}$. One may see that curves for lower clusters are identical. The curves for higher clusters, $S_{10}$ and higher, differ, but the difference is not critical. This is a favorable conclusion because the approximate model is faster to operate and so that it can be reliably used for engineering purposes.

Figure 2 shows a comparison between results of computations with the direct model run at the same conditions as those in Figure 1 but with different spans of cluster numbers included (that defines the rank of matrices in the model equations of Section 4). Results for the maximum cluster number $n_{\max} = 24$ are practically coincident with those for $n_{\max} = 36$ and most curves for them fully overlap and are not shown. The only difference is seen in the curve for higher clusters but it is very small. Differences start to be seen when $n_{\max}$ in the computations is reduced down to 12 (i.e. to $S_{24}$). This defines a restriction "at least" for cluster numbers to take into account for reliable modeling of this particular chemical process as well as contributes into understanding of a range of cluster sizes that can be formed.

At final, let us accentuate a qualitative knowledge that emerges from the present study. This is the appearance of thermal diffusion that can be significant for clusters even when transport of original chemical species in molecular forms is not a subject of such effect. It is common that vortex flows with high speed of rotation are introduced into the design with a need to centrifuge heavier substances and separate them. However, under gradients of pressure $p$ and temperature $T$ opposite to each other, the appearing thermal diffusion may prevail. Let us derive a simple criterion when this effect may dominate by considering large heavy clusters only that supposedly have to be centrifuged.

We set forth the problem as to determine a direction where a mole of heavy clusters of size $n$ move if a heat source is placed in the center of the flow rotation, and the clusters are located at radius $r$ from the center. The centrifugal particle flux $j_{n,a}^N$ out of the center is:

$$j_{n,a}^N = N x_n b_n \mu_n a \qquad (5.2)$$

where $N$ is the gas molar density, $x_n$ is mole fraction, $\mu_n = n\mu_1$ is the cluster molecular weight, $a$ is acceleration, and $b_n$ is mobilty of the clusters in a force field. The latter is connected with the diffusion coefficient $D_n$ by Einstein-Smoluchowski relation:

$$b_n = \frac{D_n}{RT}; \quad a = \frac{w^2}{r} \qquad (5.3)$$

and $w$ is the flow rotation velocity. Thus, the centrifugal flux is equal to:

$$j_{n,a}^N = N D_n x_n n \left( \mu_1 \frac{w^2}{RT} \right) \frac{1}{r} \qquad (5.4)$$

The expression in parentheses in this equation is actually a square of a ratio between $w$ and the speed of sound $c_s$ (in a gas of monomers) that usually is less than unity.

By the other hand, the thermal diffusional flux is:

$$j_{n,T}^N = -N D_n \left( \frac{dx_n}{dr} \right)_{p,x_1} \cong N D_n x_n n \nu \frac{1}{T} \frac{dT}{dr}; \quad \nu = \frac{\Delta H}{RT} \gg 1; \quad \frac{1}{T} \frac{dT}{dr} \cong -\frac{\Delta T}{T} \frac{1}{r} \qquad (5.5)$$

where we used equations (1.3) and expressed the gradient of $K$ by equation (1.10) for clusters $n \gg 1$ only. $\Delta T$ is a characteristic radial change in temperature. Flux $j_{n,T}^N$ is directed as the temperature gradient, and as the heat source is in the center of rotation, this flux is opposite to the centrifugal one. By comparing equations (5.4) and (5.5), a ratio between them is:

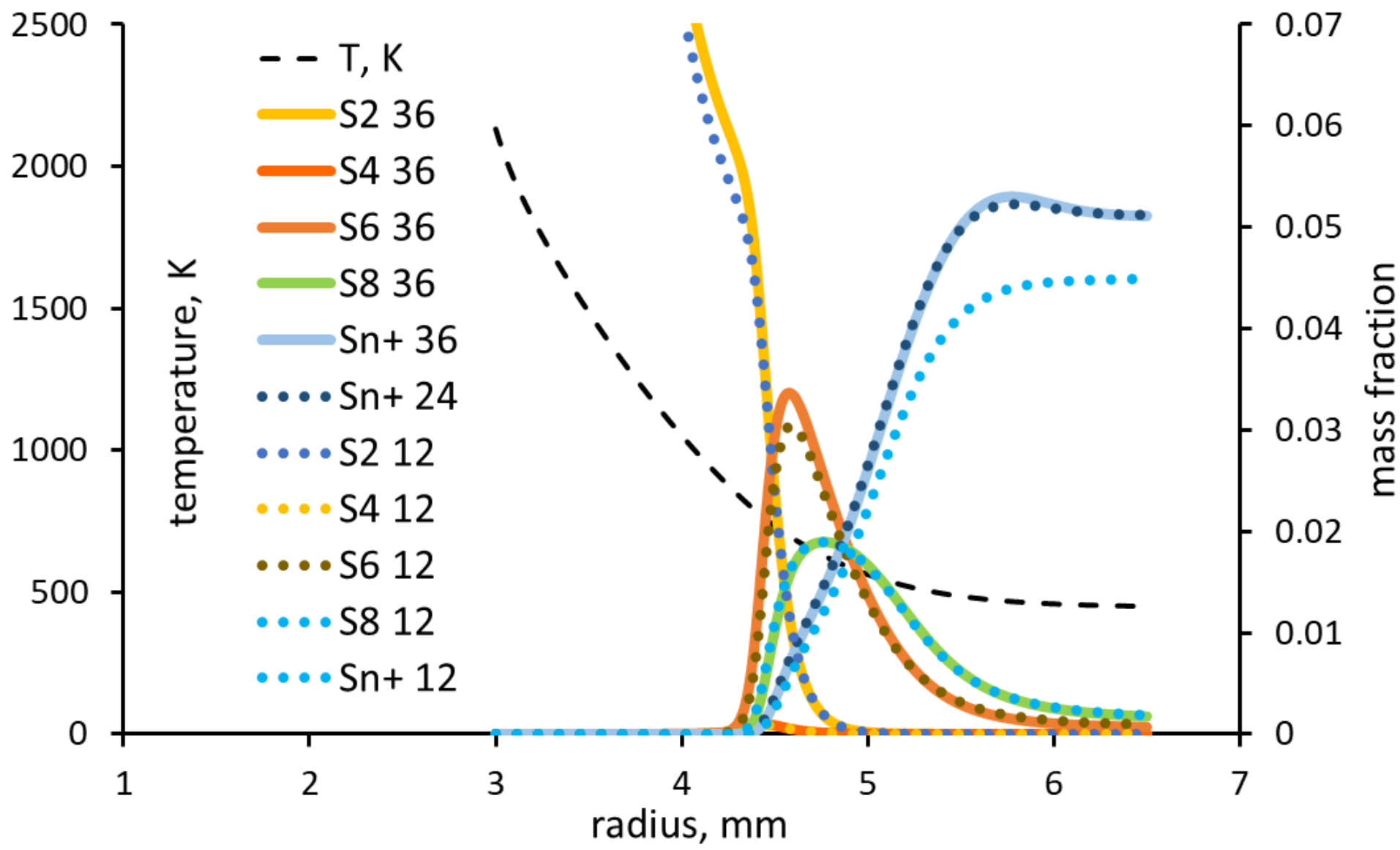

Figure 2. Comparison of spatial distributions of sulfur clusters in a centrifugal plasma-chemical reactor predicted by the direct model with different maximum cluster numbers $n_{\max}$. The $n_{\max}$ is indicated in the legend after the cluster designations. Sn+ stands for cumulative mass fraction of clusters $S_{10}$ and above. The number of S atoms in a cluster is twice the cluster number.

---

$$\left|\frac{j_{n,T}^{N}}{j_{n,a}^{N}}\right| \cong \nu \frac{\Delta T}{T} = \frac{\Delta H}{RT} \frac{\Delta T}{T} \tag{5.6}$$

even if to consider the rotational velocity $w$ at maximum, close to $c_s$. Because the heat $\Delta H$ of the cluster formation reaction (1.1) is typically much larger than $RT$, factor $\nu$ in ratio (5.6) is large. Thus, a centrifugal effect for clusters may dominate only in areas where the temperature profile is sufficiently flat or the temperature gradient is opposite. In a general case, a convection flux has to be also taken into account.

In summary, the developed theoretical approach is capable of effectively describing transport of a chemical species that forms a multitude of clusters as transport of a single species in gas. A partial chemical equilibrium between clusters is an assumption in this approach. Closed-form expressions for the effective diffusion and thermal diffusion coefficients at the temperature, pressure, and chemical composition gradients have been derived. The theory has been approbated in an application to a real chemical technological process and essentially has made it possible to account for clusters quantitatively in numerical process modeling.

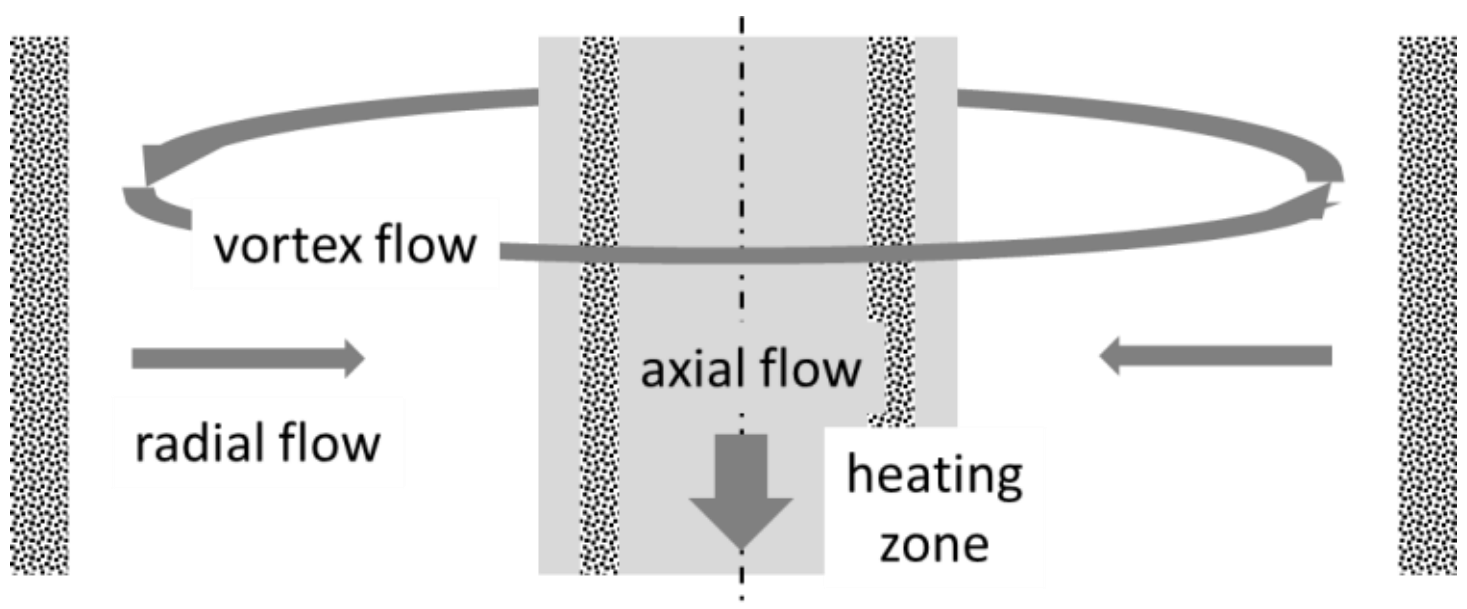

Figure 3. A schematic of the reactor 1D model geometry

---

In prospective, the theory can be extended beyond the clusters to the transport of other ensembles of chemically acting components that maintain partial equilibrium. Furthermore, when thermal diffusion of some species in gas is discovered in an experiment, this may be caused by a compound nature of the species that represents such an ensemble, and the extended theory may help to analyze the experimental data in order to elucidate the cause.

## Acknowledgment

Authors are grateful to Dr. Kirill Gutsol for sharing data from his Chemkin® modeling of the $H_2S$ thermal decomposition mechanisms. Partial financial support from NSF SBIR (grant 2233170) and Halliburton Labs Clean Energy Accelerator are highly appreciated.

## A. Appendix: Basic equations of the numerical model

**T**he model is depicted in Figure 3. In order to formulate 1D equations for major processes in the reactor, it is envisioned as two coaxial porous cylinders through which gas can be supplied or consumed. The $H_2S$ gas is supplied through the outer cylinder that rotates with high azimuthal speed, potentially close to the speed of sound. The outer cylinder rotation causes the gas inside to rotate as well. The inner cylinder rotates in the same direction. Options have been considered, when both cylinders rotate either with the same angular velocity, with given constant linear velocity, or with constant angular momentum. A concurrent radial flow with the velocity much lower than the vortex flow delivers the injected gas into the inner porous cylinder where gas leaves the reactor by an axial flow to an orifice. The area between the porous cylinders represents the computational domain. The model is stationary.

The inner cylinder is heated by the powerful plasma discharge. In a basic option, the discharge is fully contained inside so heating occurred only by heat conductance from the inner cylinder.

The heat causes the thermal decomposition of $H_2S$ by reaction (5.1). At first, an original model only for principal components of this reaction is described, and then the incorporation of the sulfur clusters is discussed.

The mass conservation for species is considered in terms of total radial mass fluxes $J_\alpha(r)$ that include both convective and diffusional fluxes. In equations below, the symbol indexes in Greek letters still correspond to species $A$, $B$, and $C$, as is in the main part of the article, which in this case denote $H_2S$, $H_2$, and $S_2$, respectively. In the radial symmetry, the equations are:

$$\frac{1}{r}\frac{d}{dr} rJ_\alpha(r) = \mu_\alpha W_\alpha(r) = \mu_\alpha \zeta_\alpha W_A(r); \quad \zeta = \{1, -1, -1/2\} \tag{A.1}$$

where $W_\alpha(r)$ are the molar volumetric rates of production of species $\alpha$ that relate to each other by stoichiometric coefficients $\zeta_\alpha$ of equation (5.1). It is convenient to express them in terms of $W_A(r)$ for $H_2S$. Furthermore, let us define functions:

$$\Phi_\alpha(r) = rJ_\alpha(r); \quad F(r) = \int_{r_0}^{r} rW_A(r)dr \tag{A.2}$$

where $r_0$ is the radius of the inner cylinder. The sum of the right sides of equations (A.1) for all species is zero, which corresponds to a conservation of the total radial mass flux $\Phi$ that is the externally injected flow of $H_2S$ (toward the center). Because flow in or out of a porous cylinder is convective, equations (A.1) become:

$$\Phi_\alpha(r) = \Phi\omega_{0,\alpha} + \mu_\alpha \zeta_\alpha F(r) \tag{A.3}$$

where $\omega_{0,\alpha}$ are the mass fractions of the species at the inner cylinder as they leave the computational domain. The boundary conditions at the radius $r_w$ of the outer cylinder for zero fluxes of the reaction products $H_2$ and $S_2$ provide connections between $\omega_{0,\alpha}$ and the total degree of the $H_2S$ decomposition $\xi$:

$$\omega_{0,A} = 1 - \xi; \quad \omega_{0,B} = \frac{\mu_B}{\mu_A}\xi; \quad \omega_{0,C} = \frac{\mu_C}{2\mu_A}\xi; \quad \xi = \mu_A \frac{F(r_w)}{\Phi} \tag{A.4}$$

which makes equations (A.3) one-parametric and dependent only upon $W_A(r)$, the conversion rate of $H_2S$, that provides increments for function $F(r)$. (For $H_2S$ decomposition, both $W_A(r)$ and $F(r)$ are negative, and the total radial mass flow $\Phi$ toward the center is negative too).

The $H_2S$ conversion (5.1) is actually a complex process that includes formation of intermediate radicals and reactions between them. These kinetics have been comprehensively explored and modeled by Chemkin® software with 9 and 16 radical reactions [51]. In the range of interest for temperatures around 2000 K and pressures above 0.1 atm, it has been possible to effectively interpolate the decomposition rate by a function that involves mole fractions $x_\alpha$ of principal components only:

$$W_A = W_A(p, T, x_A, x_B, x_C\ ) \tag{A.5}$$

where $x_C$ stands in this equation for $S_2$ (the monomer, if clusters are further formed). This interpolation facilitates the current modeling.

Analysis of the momentum conservation equations for a viscous flow of the compressible gas in the vortex shows that, in the present model, it is sufficient to consider only an equation for the pressure gradient:

$$\frac{dp}{dr} = \rho \frac{w^2}{r} \tag{A.6}$$

where $w = w(r)$ is azimuthal velocity.

The energy conservation equation is utilized in the following form:

$$\frac{1}{r}\frac{d}{dr} r \left\{ -\kappa(T)\frac{dT}{dr} + \sum\nolimits_{\alpha} \left[ h_\alpha(T) + \frac{w^2}{2} \right] J_\alpha \right\} = W_Q \tag{A.7}$$

This equation is adapted for a radial geometry and the dominance of the azimuthal flow from a multitude of forms for the energy transfer equations [28,52]. Here, $\kappa(T)$ is the thermal conductivity coefficient for the gas mixture and $W_Q$ is the volumetric heat release term. If the electric discharge is contained inside the inner porous cylinder, $W_Q = 0$. In this equation, specific enthalpies $h_\alpha(T)$ per mass of the components are full enthalpies, including the enthalpies of formation, which accounts for the chemical reaction heat in the balance.

By integrating equation (A.7) with an inner boundary condition for the heat flux $Q$ from the discharge, an equation for temperature is obtained:

$$-\kappa(T) r \frac{dT}{dr} = Q + \sum\nolimits_{\alpha} \left[ h_\alpha(T_0)\Phi_\alpha(r_0) - h_\alpha(T)\Phi_\alpha(r) \right] - \Phi \left[ \frac{w(r)^2}{2} - \frac{w_0^2}{2} \right] \tag{A.8}$$

where index 0 relates to the inner border of the computational domain. Temperature $T_0$ serves as a parameter for numerical integration from a given temperature of the cold wall (the outer porous cylinder).

The system of the balance equations is finalized by Maxwell-Stefan equations that, for principal components with no thermal diffusion, are written as follows:

$$r\frac{dx_\alpha}{dr} = -(x_\alpha - \omega_\alpha)\frac{r}{p}\frac{dp}{dr} - \mu \sum\nolimits_{\beta \neq \alpha} \frac{\omega_\beta \Phi_\alpha - \omega_\alpha \Phi_\beta}{\mu_\alpha \mu_\beta \mathcal{D}_{\alpha\beta}}; \quad \mu = \sum\nolimits_{\alpha} \mu_\alpha x_\alpha \tag{A.9}$$

The closure of the equation set (A.3)-(A.6) and (A.8)-(A.9) is provided by equation (1.4) and the equation of state for ideal gas.

Binary molecular diffusivities $\mathcal{D}_{\alpha\beta}(T)$ throughout the present study are evaluated by Fuller-Schettler-Giddings correlations [53,54]. Thermal conductivities $\kappa(T)$ are obtained by both direct data and correlations with viscosities that are usually available for broader temperature ranges [54]. For $\kappa(T)$ in a gas mixture, Wassiljewa equation [53] is used. Thermodynamic functions such as enthalpies and equilibrium constants are either obtained from NIST Chemistry Webbook [55] or generated by Chemical WorkBench® software [56].

Equations for the species concentrations (A.3)-(A.5), (A.9) and temperature (A.8) are formulated as boundary-value problems [57], and a numerical solution is achieved by applying a shooting method consecutively to converge parameters $\xi$ and $T_0$. It should be noted that a solution can be achieved in physically meaningful numbers only in limited intervals of values for the shooting parameters, and a special subroutine searches for these intervals before starting iterations. At some input conditions, a solution may not exist at all, which is not uncommon for a system of non-linear differential equations.

Incorporation of sulfur clusters into the outlined numerical model is straightforward. Equation (A.9) is replaced by equation (4.15) that is written in the 1D geometry as follows:

$$-r\frac{dx_a}{dr} = (x_\alpha - \omega_\alpha)\frac{r}{p}\frac{dp}{dr} + \mu\sum\nolimits_{\beta\neq\alpha}\frac{\omega_\beta\Phi_\alpha - \omega_\alpha\Phi_\beta}{\mu_\alpha\mu_\beta\mathcal{D}_{\alpha\beta}} + \mu\sum\nolimits_{\beta\neq\alpha}\frac{\omega_\beta D_\alpha^T - \omega_\alpha D_\beta^T}{\mu_\alpha\mu_\beta\mathcal{D}_{\alpha\beta}}\frac{r}{T}\frac{dT}{dr} \quad \text{(A.10)}$$

where binary diffusivities $\mathcal{D}_{\alpha\beta}$ and thermal diffusion coefficients $D_\alpha^T$ are evaluated by equations (4.14), (3.11), and (3.12) if the direct (matrix) theoretical formalism is used or equations listed in Table 1 for the approximate one. As "magic" clusters, $S_4$, $S_6$, and $S_8$ are considered, and $S_2$ is the cluster monomer, equations of Section 3 are used. Mixture parameters $x_C$, $\omega_C$, $\mu_C$, and $\mu$ are calculated by equations (3.1), (3.2), and (1.4). Because sulfur is still produced in reaction (5.1) as $S_2$ and only then is redistributed over clusters, equations (A.3)-(A.5) remain valid. The species concentration profiles are computed by iterations of equation (A.10), including the total sulfur mole fraction $x_C$, and the mole fraction of $S_2$ has to be extracted from $x_C$ by equations (3.2) at each consecutive radial step to be inserted into equation (A.5). Accordingly, the initial species mole fractions at the inner border of the computational domain have to be calculated from equations (A.4) with equations (3.1) and (3.2) in order to start iterating equation (A.10). Pressure is calculated concurrently. Matrix operations were performed with package Math.Net Numerics [58] available in integrated development environment Microsoft Visual Studio® [59].

The computed mass fluxes $\Phi_\alpha$ of the species are then inserted into the temperature equation (A.8) for the consecutive shooting algorithm. An account for clusters is taken by calculating the sulfur specific enthalpy $h_C$ with weight fractions of clusters. For "magic" clusters, $S_4$, $S_6$, and $S_8$, specific enthalpies are known. For large clusters $S_{10+}$, we use specific enthalpy of liquid sulfur. It may be also noted that, if thermal diffusion coefficients $D_\alpha^T$ appear in computations of mass fluxes, the enthalpy flux has to be added by so-called Dufour term [28,30]. We consider this term negligible because, by the order of magnitude, it is as the gas temperature in comparison with the total enthalpy of all species.

Additional features have been also included into the described program. The heating zone can be extended outside the inner cylinder. Sulfur condensation on the cold wall is accounted for by adding a countercurrent diffusional flux of sulfur toward the wall into the balance of fluxes (A.3) and comparing the obtained sulfur partial pressure with the saturated one. The software described above is in use by RedShift Energy, Inc. for developing and scaling up plasma-chemical reactors for $H_2S$ dissociation.

## Nomenclature

| | |
|---|---|
| $p$ | pressure, Pa |
| $p_0$ | standard pressure 1 atm, Pa |
| $T$ | temperature, K |
| $R$ | gas constant, J mol$^{-1}$ K$^{-1}$ |
| $N$ | mole gas density, mol m$^{-3}$ |
| $\rho$ | mass gas density, kg m$^{-3}$ |
| $G$ | Gibbs free energy, J mol$^{-1}$ |
| $H$ | enthalpy, J mol$^{-1}$ |
| $h$ | specific enthalpy per mass, J kg$^{-1}$ |
| $\nu$ | ratio of reaction heat to temperature |
| $K_p$ | equilibrium constant for partial pressures |
| $K$ | equilibrium constant for mole fractions |
| $x$ | mole fraction |
| $\omega$ | weight fraction |
| $\mu$ | molecular weight, kg mol$^{-1}$ |
| $\boldsymbol{d}$ | diffusional driving force, m$^{-1}$ |
| $\boldsymbol{j}$ | mass flux, kg m$^{-2}$ s$^{-1}$ |
| $j^N$ | particle flux, mol m$^{-2}$ s$^{-1}$ |
| $D$ | diffusion coefficient, binary if indexed, m$^2$ s$^{-1}$ |
| $D^T$ | thermal diffusion coefficient, kg m$^{-1}$ s$^{-1}$ |
| $\mathcal{D}$ | reduced diffusion coefficient, binary if indexed, mol m$^{-1}$ s$^{-1}$ |
| $\tilde{D}$ | generalized Fick diffusion coefficient, m$^2$ s$^{-1}$ |
| $\tilde{\mathcal{D}}$ | reduced generalized Fick diffusion coefficient, mol m$^{-1}$ s$^{-1}$ |
| $\Theta_C$ | a term defined by Eq. (3.6) |
| $\Lambda$ | matrix defined by Eq. (4.3), m s kg$^{-1}$ |
| $\Psi$ | matrix defined by Eqs. (4.11), m s mol$^{-1}$ |
| $Y$ | matrix defined by Eqs. (4.11) |
| $\Omega$ | matrix defined by Eqs. (4.11) |
| $\Psi^0$ | matrix defined by Eq. (4.12), m s mol$^{-1}$ |
| $b$ | mobility in a force field, mol s kg$^{-1}$ |
| $a$ | acceleration, m s$^{-2}$ |
| $c_s$ | speed of sound, m s$^{-1}$ |
| $z$ | terms defined in Appendix A |
| $J$ | total radial mass flux, diffusional plus convectional, kg m$^{-2}$ s$^{-1}$ |
| $\Phi$ | total radial mass flux into an azimuthal angle, kg m$^{-1}$ s$^{-1}$ |
| $r$ | radius, m |
| $\zeta$ | chemical reaction stoichiometric coefficient |
| $W$ | chemical reaction rate, mol m$^{-3}$ s$^{-1}$ |
| $F$ | mole conversion function, mol m$^{-1}$ s$^{-1}$ |
| $\xi$ | overall mole conversion degree |
| $w$ | azimuthal vortex velocity, m s$^{-1}$ |
| $\kappa$ | thermal conductivity coefficient, W m$^{-1}$ K$^{-1}$ |
| $Q$ | radial heat flux into an azimuthal angle, W m$^{-1}$ |
| $W_Q$ | volumetric heat release, W m$^{-3}$ |
| | **Subscripts and Indices** |
| *A, B* | molecular species in a gas mixture |
| *C* | a species that can form clusters |
| $\alpha, \beta$ | indices for species |
| $n, m$ | indices for clusters, cluster numbers |
| $S, P$ | diffusing molecular species $S$ in a buffer gas $P$ |
| $p, T, a$ | pressure, temperature, acceleration; contextual |
| $0$ | inner border, Appendix B |
| $i, j, k$ | general purpose indices, contextual |